\documentclass[12pt]{iopart}

\usepackage{iopams}  

\usepackage{cite}
\usepackage{graphicx}
\usepackage{bm}

\usepackage{color}

\begin{document}

\title[Combinatorics for calculating expectation values of functions...]{Combinatorics for calculating expectation values of functions in systems with evolution governed by stochastic differential equations}

\author{Jun Ohkubo$^{1,2}$}

\address{$^1$ Graduate School of Science and Engineering, Saitama University, \\
255 Shimo-Okubo, Sakura-ku, Saitama 338-8570, Japan}
\address{$^2$ JST, PRESTO, 4-1-8 Honcho, Kawaguchi, Saitama 332-0012, Japan}
\ead{johkubo@mail.saitama-u.ac.jp}
\vspace{10pt}
\begin{indented}
\item[]
\end{indented}

\begin{abstract}
{
Stochastic differential equations are widely used in various fields; in particular, the usefulness of duality relations has been demonstrated in some models such as population models and Brownian momentum processes. In this study, a discussion based on combinatorics is made and applied to calculate the expectation values of functions in systems in which evolution is governed by stochastic differential equations. Starting with the duality theory of stochastic processes, some modifications to the interpretation and usage of time-ordering operators naturally lead to discussions on combinatorics. For demonstration, the first and second moments of the Ornstein--Uhlenbeck process are re-derived from the discussion on combinatorics. Furthermore, two numerical methods for practical applications are proposed. One method is based on a conventional exponential expansion and the Pad{\'e} approximation. The other uses a resolvent of a time-evolution operator, along with the application of the Aitken series acceleration method. Both methods yield reasonable approximations. Particularly, the resolvent and Aitken acceleration show satisfactory results. These findings will provide a new way of calculating expectations numerically and directly without using time-discretization.
}
\end{abstract}

%
%
%
%
%

\section{Introduction}
\label{sec_introduction}

Stochastic differential equations, or Langevin equations, are widely used in various research fields \cite{Gardiner_book}.
The paths of stochastic differential equations are discussed based on Ito calculus, and the corresponding Fokker--Planck equations directly deal with probability density functions.
However, because obtaining analytical solutions is generally difficult, Monte Carlo simulations are employed practically.
Using numerical methods such as the Euler--Maruyama approximations, 
probability density functions are estimated adequately \cite{Kloeden_book}.

In practical cases, only limited statistics, such as averages and variances, are usually required.
Recent studies on duality relations in stochastic processes have revealed that such statistical quantities in stochastic differential equations, especially moments, can be evaluated by the corresponding dual birth--death processes as well \cite{Liggett_book,Jansen2014}.
In some cases, analytical solutions for the corresponding dual birth--death processes are available, and the usefulness of the duality relations has been demonstrated for population models \cite{Shiga1986,Mohle1999,Carinci2015} and Brownian momentum processes \cite{Giardina2007,Carinci2013}.
Recently, many studies have been conducted on duality relations from a mathematical perspective \cite{Liggett_book,Giardina2009,Franceschini2017,Redig2018,Groenevelt2019}. In this study, we focus on the numerical applications of duality relations to obtain statistical quantities from analytically intractable stochastic differential equations.
Although some applications are found with the aid of numerical calculations \cite{Ohkubo2015}, such numerical studies are still limited.
A problem with the practical usage of duality relations is the stochasticity in the dual processes.
For example, in an application to a filtering problem, large sample sizes are required to construct practical filtering procedures \cite{Ohkubo2015}.

In this study, a natural connection is established between the conventional duality relations in stochastic processes and combinatorial frameworks.
The use of combinatorics avoids the stochasticity in dual processes and is advantageous for low-dimensional cases.
Such discussions on combinatorics have recently been used to compute the Mori--Zwanzig memory integral in generalized Langevin equations \cite{Amati2019, Zhu2020}.
It was revealed that combinatorial algorithms are beneficial for evaluating operator exponentials, with the recursive algorithm efficiently computing the expansion coefficients. 
Some aspects of these recent developments apply to the construction of our algorithm for evaluating statistical quantities in stochastic differential equations. 
In this work, discussions are made related to combinatorics, and two candidates for practical numerical techniques to evaluate statistical quantities are proposed.
One candidate is based on a conventional exponential expansion and the Pad{\'e} approximation and the other uses a resolvent of a time-evolution operator and the Aitken series-acceleration method. 
Furthermore, numerical comparisons with Monte Carlo simulations are presented.

Here, to avoid any misunderstandings of the aim of this study, some notifications are given: The aim is to calculate expectation values at a certain time, not to propose time-integration methods. Hence, there is no need to apply the proposed method repeatedly to the next numerical integration. Calculating statistics over a certain time interval is sufficient for, for example, a non-linear Kalman filter. One of the possible targets of this manuscript is the calculation of expectations in such data analysis methods. For these targets, it is often sufficient to know only a few statistics. The method proposed here is fit for the purpose, as it computes the statistics without directly evaluating the probability density function.

The remainder of this paper is organized as follows.
In section 2, previous works are briefly reviewed, and the first and second moments of the Ornstein--Uhlenbeck process are re-derived from the duality relations in stochastic processes.
Section 3 focuses on the connection between duality relations and combinatorics. Further, the interpretation of the dual birth--death process is modified, and the time-ordering operator is used to make the connection. Additionally, discussions on combinatorics are demonstrated using the Ornstein--Uhlenbeck process.
Section 4 discusses an analytically intractable example and proposes two numerical methods. 
Concluding remarks are presented in section 5.

\section{Brief review of previous works}
\label{sec_dual_review}

\subsection{Duality relations between stochastic differential equations and birth--death processes}

Although discussions on combinatorics can be made without using duality relations, it is beneficial to clarify their connections.
Thus, the duality in stochastic processes is briefly reviewed here.

As stated in the Introduction, the duality relations in stochastic processes are widely used in various fields, including interacting particle systems such as simple exclusion processes \cite{Schutz1997,Imamura2011,Ohkubo2017}.
In this study, we focus on the duality relations between stochastic differential equations and the birth--death process. 
Here, stochastic differential equations with only one variable are discussed; the following discussions can be straightforwardly applied to multivariate cases.

Let $x_t \in \mathbb{R}$ be the state of a stochastic differential equation at time $t$.
The corresponding dual birth--death process is a stochastic process with discrete state and continuous time, and its state at time $t$ is written as a state vector $\bm{n}_t \in \mathbb{N}^{d_\mathrm{dual}}$, where $d_\mathrm{dual}$ denotes the number of variables in the dual process.
Note that these two processes need not have the same dimensions (discussed later).
Process $(x_t)$ is said to be dual to $(\bm{n}_t)$ with respect to a duality function $D: \mathbb{R} \times \mathbb{N}^{d_\mathrm{dual}} \to \mathbb{R}$, if for all $(x_t)$, $(\bm{n}_t)$, and $t \ge 0$ we have
\begin{eqnarray}
\mathbb{E}_{\bm{n}_0} \left[ D(x_0,\bm{n}_t)\right]
= \mathbb{E}_{x_0} \left[ D(x_t,\bm{n}_0)\right],
\label{eq_basic_duality}
\end{eqnarray}
where $\mathbb{E}_{x_0}$ and $\mathbb{E}_{\bm{n}_0}$ are the expectations in the processes $(x_t)$ starting from $x_0$ and $(\bm{n}_t)$ starting from $\bm{n}_0$, respectively.
This duality relation indicates that the solution of the dual birth--death process gives the expectations in the original stochastic differential equation.

Therefore, deriving the dual birth--death process is the remaining problem. 
In the following sections, this derivation is briefly reviewed and demonstrated using the Ornstein--Uhlenbeck process.

\subsection{Kolmogorov backward equation and dual process}
\label{sec_dual_process}

We start with the following stochastic differential equation:
\begin{eqnarray}
\rmd x = \mu(x) \rmd t + \sigma(x) \rmd W(t),
\end{eqnarray}
where $\mu(x)$ is the drift coefficient, $\sigma(x)$ is the diffusion coefficient, and $W(t)$ represents the Wiener process.
Although the coefficients can be time-dependent, only time-independent cases are considered in this study.
The stochastic differential equation has the following corresponding partial differential equation, i.e., Fokker--Planck equation \cite{Gardiner_book}:
\begin{eqnarray}
\frac{\partial}{\partial t} p(x,t) = \mathcal{L} p(x,t),
\end{eqnarray}
where $p(x,t)$ is a probability density function for $x$ at time $t$, and
\begin{eqnarray}
\mathcal{L} \equiv - \frac{\partial}{\partial x} \mu(x)
+ \frac{1}{2}  \frac{\partial^2}{\partial x^2} \sigma(x)^2
\end{eqnarray}
is a time-evolution operator.

In contrast to the Fokker--Planck equation, it is known that the Kolmogorov backward equation (or backward Fokker--Planck equation) starts from a final condition and is integrated backward in time \cite{Gardiner_book}.
The famous Feynman--Kac formula extends the discussion of the Kolmogorov backward equation, and these are starting points of the derivation of duality relations in stochastic processes.
As discussed in \cite{Ohkubo2019}, the derivation of the corresponding dual birth--death process can be comprehended only by the use of integration-by-parts and function expansions.
When we consider the $m$th moment of $x$ in the stochastic differential equation, it is possible to formally rewrite it as follows:
\begin{eqnarray}
\mathbb{E}\left[ x^m \right]
&= \int_{-\infty}^{\infty} x^m p(x,T) \rmd x \nonumber \\
&= \int_{-\infty}^{\infty} x^m \left( \rme^{\mathcal{L} T} \delta(x-x_0) \right) \rmd x \nonumber \\
&= \int_{-\infty}^{\infty} \left( \rme^{\mathcal{L}^\dagger T} x^m  \right)  \delta(x-x_0) \rmd x \nonumber \\
&= \int_{-\infty}^{\infty} \widetilde{\varphi}(x,T) \delta(x-x_0) \rmd x \nonumber \\
&= \widetilde{\varphi}(x_0,T),
\end{eqnarray}
where $\delta(x)$ is the Dirac delta function, $x_0$ is the initial position, and $T$ is the final time at which we evaluate the expectation.
Note that the function $\widetilde{\varphi}$ is not a probability function; the time-evolution operator for $\widetilde{\varphi}$ is given by
\begin{eqnarray}
\mathcal{L}^\dagger \equiv \mu(x) \frac{\partial}{\partial x}
+ \frac{1}{2} \sigma(x)^2 \frac{\partial^2}{\partial x^2},
\label{eq_adjoint_operator}
\end{eqnarray}
which is the adjoint operator of $\mathcal{L}$ and does not satisfy the probability conservation law in general.

The time-evolution equation for $\widetilde{\varphi}$,
\begin{eqnarray}
\frac{\partial}{\partial t} \widetilde{\varphi}(x,t) 
= \mathcal{L}^\dagger \widetilde{\varphi}(x,t),
\end{eqnarray}
should be performed in backward from $t=T$ to $t=0$.
This backward evolution could sometimes be confusing if $\mathcal{L}$ depends on time $t$.
To avoid this confusion, we here rewrite $t$ as $T - t$ and define 
$\varphi(x,t) \equiv \widetilde{\varphi}(x,T-t)$; the function $\varphi(x,t)$ is integrated forward in time from $t = 0$ to $t = T$.
Note that the time-evolution has the following initial condition:
\begin{eqnarray}
\varphi(x,0) = x^m.
\end{eqnarray}

The function $\varphi(x,t)$ still has a continuous variable $x$, and the dual birth--death process is derived using function expansions.
Various expansions are available for this purpose; Hermite polynomials \cite{Ohkubo2019} and Legendre polynomials \cite{Ohkubo2020} have been used in previous works (for details, see \cite{Ohkubo2019}.)
To aid the understanding of the discussion, a demonstration of the derivation of the dual birth--death process using the Ornstein--Uhlenbeck process is presented in the next subsection.

\subsection{Example: Ornstein--Uhlenbeck process}
\label{sec_dual_demonstration}

The famous Ornstein--Uhlenbeck process is solved analytically \cite{Gardiner_book}:
\begin{eqnarray}
\rmd x = - \gamma x \rmd t + \sigma \rmd  W(t),
\end{eqnarray}
where $\gamma > 0$ and $\sigma > 0$.
The adjoint time-evolution operator $\mathcal{L}^\dagger$ is given as
\begin{eqnarray}
\mathcal{L}^\dagger = - \gamma x \frac{\partial}{\partial x} 
+ \frac{\sigma^2}{2} \frac{\partial^2}{\partial x^2}.
\end{eqnarray}
Here, for later use, we introduce the coordinate transformation $x' = x - x_\mathrm{c}$; rewriting $x'$ as $x$ again, the adjoint operator is rewritten as
\begin{eqnarray}
\mathcal{L}^\dagger &= - \gamma (x+x_\mathrm{c}) \frac{\partial}{\partial x} 
+ \frac{\sigma^2}{2} \frac{\partial^2}{\partial x^2} \nonumber \\
&= - \gamma x  \frac{\partial}{\partial x} 
- \gamma x_\mathrm{c} \frac{\partial}{\partial x} 
+ \frac{\sigma^2}{2} \frac{\partial^2}{\partial x^2}.
\label{eq_OU_adjoint_operator}
\end{eqnarray}
This coordinate transformation is introduced to calculate the Taylor expansion around $x_\mathrm{c}$.
When $x_\mathrm{c}$ is the initial condition of the original stochastic differential equation, it is sufficient to consider only $\varphi(0,T)$ for evaluating the expectation values.

The following expectation values are obtained using the established analytical solution of the Ornstein--Uhlenbeck process \cite{Gardiner_book}:
\begin{eqnarray}
\mathbb{E}\left[ x_T - x_\mathrm{c}\right] = x_\mathrm{c} \rme^{-\gamma T} - x_\mathrm{c},
\label{eq_OU_1st_moment}
\end{eqnarray}
\begin{eqnarray}
\mathbb{E}\left[ (x_T - x_\mathrm{c})^2 \right] 
= \frac{\sigma^2}{2\gamma} \left( 1- \rme^{-2\gamma T}\right)
+ x_\mathrm{c}^2 \rme^{-2\gamma T}
- 2 x_\mathrm{c}^2 \rme^{-\gamma T}
+ x_\mathrm{c}^2.
\label{eq_OU_2nd_moment}
\end{eqnarray}
Next, we demonstrate that the dual birth--death process can recover these two expressions.
To recover the stochasticity of the dual process, the adjoint operator $\mathcal{L}^\dagger$ for the time-evolution is split into two parts:
\begin{eqnarray}
\mathcal{L}^\dagger = \widetilde{\mathcal{L}}^\dagger + V,
\end{eqnarray}
where
\begin{eqnarray}
\widetilde{\mathcal{L}}^\dagger 
= \gamma x_\mathrm{c} x^{(0)} \frac{\partial}{\partial x} - \gamma x_\mathrm{c} x \frac{\partial}{\partial x} 
+ \frac{\sigma^2}{2} \frac{\partial^2}{\partial x^2} - \frac{\sigma^2}{2} x^2 \frac{\partial^2}{\partial x^2}, 
\label{eq_OU_avoid_negative}\\
V = - \gamma x  \frac{\partial}{\partial x} + \gamma x_\mathrm{c} x \frac{\partial}{\partial x} 
+ \frac{\sigma^2}{2} x^2 \frac{\partial^2}{\partial x^2},
\label{eq_OU_Feynman_Kac}
\end{eqnarray}
and $x^{(0)} \equiv -1$ is introduced to avoid a negative transition problem; this variable is important to form a corresponding stochastic process \cite{Ohkubo2013}. The role of the variable $x^{(0)}$ will be explained later.
It is easy to see that the operator $\widetilde{\mathcal{L}}^\dagger$ gives the time-evolution for a dual birth--death process, and $V$ corresponds to the Feynman--Kac term.
Using the function expansion
\begin{eqnarray}
\varphi(x,t) = \sum_{n=0}^\infty P(n, n_0, t) (x^{(0)})^{n_0} x^n,
\end{eqnarray}
the time-evolution equation
\begin{eqnarray}
\frac{\partial}{\partial t} \varphi(x,t) = \widetilde{\mathcal{L}}^\dagger \varphi(x,t)
\end{eqnarray}
gives the following equation for the coefficients $P(n,n_0,t)$:
\begin{eqnarray}
\fl
\frac{\rmd}{\rmd t} P(n, n_0, t) 
= & \gamma x_\mathrm{c} (n+1) P(n+1, n_0-1, t) - \gamma x_\mathrm{c} n P(n,n_0, t) \nonumber \\
\fl
&+ \frac{\sigma^2}{2} (n+2)(n+1) P(n+2, n_0, t) - \frac{\sigma^2}{2} n(n-1) P(n, n_0, t).
\label{eq_OU_master_equation}
\end{eqnarray}
This master equation for $P(n,n_0,t)$ is interpreted as the following chemical reaction system:
\begin{eqnarray}
\begin{array}{ll}
\textrm{Event 1: } \,\, X \to X^{(0)} \quad & \textrm{at rate $\gamma x_\mathrm{c} n$},\\
\textrm{Event 2: } \,\, X + X \to \varnothing & \textrm{at rate $\sigma^2 n(n-1)/2$}.
\end{array}
\label{eq_OU_chemical_reactions}
\end{eqnarray}

If the variable $x^{(0)}$ is not introduced, the first and second terms in \eref{eq_OU_avoid_negative} give
\begin{eqnarray*}
- \gamma x_\mathrm{c} \frac{\partial}{\partial x} - \gamma x_\mathrm{c} x \frac{\partial}{\partial x}.
\end{eqnarray*} 
These terms yield an equation that cannot be simply interpreted as a master equation for a stochastic process, i.e., the first term leads to a rate constant $-\gamma x_\mathrm{c} (n+1)$, which has a negative sign.
The new variable $x^{(0)}$ adequately yields the stochastic process in \eref{eq_OU_chemical_reactions}.
Although the introduction of $x^{(0)}$ requires a new stochastic variable, $n_0$, in the master equation,
it enables interpretation in the form of stochastic processes.

Note that the initial condition of $P(n,n_0,t)$ is $P(n=m, n_0=0,t=0) = 1$, and zero otherwise, to evaluate $\mathbb{E}[x^m]$.
The Feynman--Kac term in \eref{eq_OU_Feynman_Kac} should also be considered; it can be interpreted in terms of the variables of the dual birth--death process as follows:
\begin{eqnarray}
V(n) = - \gamma n + \gamma x_\mathrm{c} n + \frac{\sigma^2}{2} n(n-1).
\end{eqnarray}
Note that this term does not depend on $n_0$.

Using the above dual birth--death process, the statistics in the original stochastic differential equation can be straightforwardly evaluated.
When the initial condition is set as $n = 1$, only Event 1 in \eref{eq_OU_chemical_reactions} is allowed; thus, we have
\begin{eqnarray}
\int_0^T \rme^{V(n=1) t_1} 
\left[ \rme^{-\gamma x_\mathrm{c} t_1} \gamma x_\mathrm{c} \right] \rmd t_1
= - \left(
x_\mathrm{c} \rme^{-\gamma T} - x_\mathrm{c}
\right),
\label{eq_OU_dual_event_1}
\end{eqnarray}
where $\left[\,\cdot\, \right]$ corresponds to the probability density for the path with only one Event 1 \cite{Gillespie1977}.
Note that Event 1 also causes the change $n_0 = 0 \to 1$.
Using the fact that $n_0 = 1$ at the final time $T$, the negative sign in \eref{eq_OU_dual_event_1} is cancelled, and the expectation value in \eref{eq_OU_1st_moment} is adequately recovered.
For $n=2$ at $t = 0$, there are two possible paths in which
\begin{enumerate}
\item[(i)] Event 2 occurs once,
\item[(ii)] Event 1 occurs twice.
\end{enumerate}
Path (i) gives
\begin{eqnarray}
\int_0^T \rme^{V(n=2) t_1} 
\left[ \rme^{- (2\gamma x_\mathrm{c} + \sigma^2) t_1} \sigma^2  \right] \rmd t_1
= \frac{\sigma^2}{2\gamma} \left(1-\rme^{-2\gamma T} \right),
\end{eqnarray}
and path (ii) provides the following contribution:
\begin{eqnarray}
\int_0^T \rmd t_2 \int_{0}^{t_2} \rmd t_1
\rme^{V(n=2) t_1} 
\left[ \rme^{- (2\gamma x_\mathrm{c} + \sigma^2) t_1} 2\gamma x_\mathrm{c}   \right] 
\rme^{V(n=1) (t_2-t_1)} 
\left[ \rme^{- (\gamma x_\mathrm{c}) (t_2-t_1)} \gamma x_\mathrm{c}   \right]  \nonumber \\
= x_\mathrm{c} \rme^{- 2 \gamma T} - 2 x_\mathrm{c}^2 \rme^{- \gamma T} + x_\mathrm{c}^2.
\label{eq_dual_calculation_for_second_moment}
\end{eqnarray}
Noting that $n_0 = 0$ for path (i) and $n_0 = 2$ for path (ii),
the second moment in \eref{eq_OU_2nd_moment} is recovered.

\section{Combinatorics}

This section presents the first main contribution of this study, namely, the discussion on combinatorics, starting from the dual birth--death process. 
The discussion is applied to the Ornstein--Uhlenbeck process, and we confirm that the analytical solutions are recovered adequately.

\subsection{From dual birth--death process to simple combinatorics}

As discussed in \cite{Ohkubo2013}, the time-evolution with $\widetilde{\mathcal{L}}^\dagger$ acts on the state vector $|\bm{n}\rangle$ as follows:
\begin{eqnarray}
\rme^{\widetilde{\mathcal{L}}^\dagger \Delta t} | \bm{n} \rangle
&\simeq 
\left(1+\widetilde{\mathcal{L}}^\dagger \Delta t \right) | \bm{n} \rangle \nonumber \\
&= 1 - \sum_{r=1}^R  a_r(\bm{n}) \Delta t | \bm{n} \rangle \langle \bm{n} | 
+ \sum_{r=1}^R  a_r(\bm{n})   \Delta t | \bm{n} + \bm{v}_r \rangle \langle \bm{n} | \nonumber \\
&\simeq \rme^{-a_0(\bm{n}) \Delta t} | \bm{n} \rangle \langle \bm{n} |
+ \left( \sum_{r=1}^R  a_r(\bm{n})  \Delta t | \bm{n} + \bm{v}_r \rangle \langle \bm{n} | \right),
\label{eq_dual_for_interpretation}
\end{eqnarray}
where $R$ denotes the number of events, $a_r(\bm{n})$ is the propensity function for the $r$th event \cite{Wilkinson_book}, and $a_0(\bm{n}) = \sum_{r=1}^R a_r(\bm{n})$;
the vector $\bm{v}_r$ represents the stoichiometric coefficients of event $r$ \cite{Warne2019}.
Although the state vector $|\bm{n}\rangle$ could be an abstract one that satisfies $\langle \bm{m} | \bm{n} \rangle = \delta_{\bm{m},\bm{n}}$, it is possible to define it as an infinite-dimensional one or an explicit one in terms of $x$ \cite{Ohkubo2010,Weber2017}.
For the example of the Ornstein--Uhlenbeck process, $R$ is $2$, and $a_1(\bm{n}) = \gamma x_\mathrm{c} n$, $a_2(\bm{n}) = (\sigma^2/2) n(n-1)$.
The comparison with \eref{eq_OU_master_equation} will make it easier to understand.

The Taylor expansion in \eref{eq_dual_for_interpretation} gives a natural probabilistic interpretation of the birth--death process.
The first and second terms in the third line in \eref{eq_dual_for_interpretation} are considered as a Bernoulli trial: with the probability $\rme^{-a_0(\bm{n}) \Delta t}$, no event occurs, whereas some events occur with the probability $a_0(\bm{n}) = \sum_{r=1}^R  a_r(\bm{n})$.
For the latter case, only an event with the probability $a_r(\bm{n}) / a_0(\bm{n})$ is chosen; the factor $1/a_0(\bm{n})$ is compensated by considering $a_0(\bm{n}) \exp\left(-a_0(\bm{n}) \Delta t\right)$, which gives the exponential distribution for the event-interval time.

For a discussion on combinatorics, the interpretation of the time-evolution operator should be slightly modified. 
That is, instead of the time-evolution operator with the probability conservation law, $\widetilde{\mathcal{L}}^\dagger$, the adjoint operator $\mathcal{L}^\dagger$ in \eref{eq_adjoint_operator} is directly considered here.
Thus, we have
\begin{eqnarray}
\rme^{\mathcal{L}^\dagger \Delta t} | \bm{n} \rangle
&\simeq 
1 + \sum_{r=1}^R  a_j(\bm{n})  \Delta t | \bm{n} + \bm{v}_r \rangle \langle \bm{n} |.
\label{eq_adjoint_expansion_reinterpretation}
\end{eqnarray}
Repeated actions of $\exp\left( \mathcal{L}^\dagger \Delta t \right)$ result in simple products of $a_j$ as follows:
\begin{eqnarray}
\rme^{\mathcal{L}^\dagger T} | \bm{n}_{\mathrm{ini}} \rangle \nonumber \\
\simeq  
a_{j^{(M)}}(\bm{n}_{M-1}) a_{j^{(M-1)}}(\bm{n}_{M-2})\cdots a_{j^{(1)}}(\bm{n}_{\mathrm{ini}})  
\left(\Delta t \right)^M \left| \bm{n}_{\mathrm{ini}} + \sum_{m=1}^M \bm{v}_{j^{(m)}}  \right\rangle,
\label{eq_time_evolution_re_interpretation}
\end{eqnarray}
where $M$ represents the number of times the events in the second term are selected in \eref{eq_adjoint_expansion_reinterpretation}, and $a_{j^{(m)}}(\bm{n})$ is the propensity function for event $j^{(m)}$.
Here, the initial state is written as $\bm{n}_\mathrm{ini} \equiv \bm{n}_1$.

Note that \eref{eq_time_evolution_re_interpretation} provides the path contribution; from the perspective of the path integral, the time-integration for all possible paths is required.
Note that the factors $\{a_{j^{(m)}}\}$ do not commute each other; for example, 
\begin{eqnarray}
a_{j^{(M)}}(\bm{n}_{M-1}) a_{j^{(M-1)}}(\bm{n}_{M-2})
\neq 
a_{j^{(M)}}(\bm{n}_{M-2}) a_{j^{(M-1)}}(\bm{n}_{M-1}),
\end{eqnarray}
because the factors depend on the state.
Therefore, we introduce the following time-ordering operator \cite{Kleinert_book}:
\begin{eqnarray}
\mathcal{T}\left\{ a_{j^{(2)}}(\bm{n}_2) a_{j^{(1)}}(\bm{n}_1)\right\}
\nonumber \\
\equiv a_{j^{(2)}}(\bm{n}_2) a_{j^{(1)}}(\bm{n}_1) \theta(t_2-t_1) 
+ a_{j^{(2)}}(\bm{n}_1) a_{j^{(1)}}(\bm{n}_2) \theta(t_1-t_2),
\end{eqnarray}
where $\theta(t)$ is the Heaviside function.
Using the abbreviation $a^{(m)} \equiv a_{j^{(m)}}(\bm{n}_{m-1})$, the probability of events occurring for $M$ times is written as follows:
\begin{eqnarray}
\int_0^T \rmd t_{M} \int_0^{t_M} \rmd t_{M-1} \cdots \int_0^{t_2} \rmd t_1 a^{(M)} a^{(M-1)} \cdots a^{(1)}
\nonumber \\
= \frac{1}{M!} \int_0^T \rmd t_{M} \int_0^T \rmd t_{M-1} \cdots \int_0^T \rmd t_1 
\mathcal{T} \left\{  a^{(M)} a^{(M-1)} \cdots a^{(1)} \right\}
\nonumber \\
= \frac{T^M}{M!}   a^{(M)} a^{(M-1)} \cdots a^{(1)}.
\label{eq_combinatrics_basic}
\end{eqnarray}

The final expression in \eref{eq_combinatrics_basic} provides the basis for the combinatorics to calculate the expectations values.
Recall that each event changes the state $\bm{n}$ in the dual (non-stochastic) process. 
Additionally, unlike the dual birth--death process in section~\ref{sec_dual_review}, the term giving the Feynman--Kac contribution is also considered as an event in the present discussion.
Note that the usage of the event with no state change is different from that of the Extrande algorithm \cite{Voliotis2016} or uniformization techniques \cite{Beentjes2019}, which are proposed for time-inhomogeneous birth--death processes.
Here, the propensity functions do not depend on time.

The general discussion can be understood clearly if we can see the exactly solvable cases of the Ornstein--Uhlenbeck process.
The time-evolution operator $\mathcal{L}^\dagger$ for the dual process is given in \eref{eq_OU_adjoint_operator}, and it acts on $|n\rangle \equiv x^n$ as follows:
\begin{eqnarray}
\mathcal{L}^\dagger  | n \rangle
= -\gamma n | n \rangle 
- \gamma x_\mathrm{c} n | n-1 \rangle 
+ \frac{\sigma^2}{2} n(n-1) | n-2 \rangle.
\end{eqnarray}
That is, there are three events as follows:
\begin{enumerate}
\item[(I)] (from $-\gamma x \partial_x$) $n \to n$ (no state change), with a factor $-\gamma n$.
\item[(II)] (from $-\gamma x_\mathrm{c} \partial_x$) $n \to n-1$, with a factor $-\gamma x_\mathrm{c} n$.
\item[(III)] (from $(\sigma^2/2) \partial^2_x$) $n \to n-2$, with a factor $\sigma^2 n(n-1)/2$.
\end{enumerate}

\subsection{First-order moment in Ornstein--Uhlenbeck process}

\begin{figure}
\begin{center}
\vskip 5mm
\includegraphics[width=80mm]{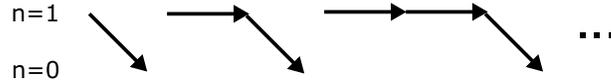}
\end{center}
\caption{Possible combinations of paths starting from $n=1$. Once the state with $n=0$ is reached, there is no further occurrence of events.} 
\label{fig_OU_combinatorics_first}
\end{figure}

In the calculation of the first-order moment $\mathbb{E}\left[ x_T - x_\mathrm{c}\right]$, the initial state is set to $n=1$.
In this case, Event 3 never occurs.
Additionally, if Event 2 occurs once, all three events are no longer permitted.
Therefore, as shown in Figure~\ref{fig_OU_combinatorics_first}, the possible combinations of paths give
\begin{eqnarray}
&\frac{T}{1!} (-\gamma x_\mathrm{c})
+ \frac{T^2}{2!} (-\gamma) (-\gamma x_\mathrm{c})
+ \frac{T^3}{3!} (-\gamma)^2 (-\gamma x_\mathrm{c}) \cdots \nonumber \\
&= x_\mathrm{c} 
\left[ 1 +\frac{T}{1!}(-\gamma) + \frac{T^2}{2!}(-\gamma)^2 + \frac{T^3}{3!}(-\gamma)^3 +  \cdots \right] 
- x_\mathrm{c}
\nonumber \\
&= x_c \rme^{-\gamma T} - x_\mathrm{c},
\end{eqnarray}
which adequately give the same analytical result as in \eref{eq_OU_1st_moment}.

\subsection{Second-order moment in Ornstein--Uhlenbeck process}

\begin{figure}
\begin{center}
\vskip 5mm
\includegraphics[width=120mm]{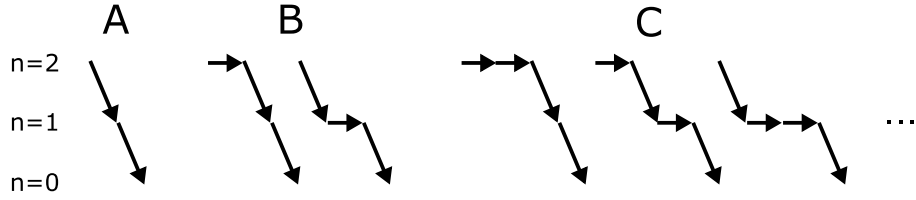}
\end{center}
\caption{Possible combinations of paths, starting from $n=2$ and using only Events 1 and 2 in the main text. A, B, and C correspond to cases with $M=1$, $M=2$, and $M=3$, respectively.}
\label{fig_OU_combinatorics_second}
\end{figure}

When the second-order moment $\mathbb{E}\left[ (x_T - x_\mathrm{c})^2 \right]$ is evaluated, the dual process starts from $n=2$.
In contrast to the previous case starting from $n=1$, complicated discussions are required.

First, Event 3 causes a sudden change from $n=2$ to $n=0$; in this case, similar to that in Figure~\ref{fig_OU_combinatorics_first}, we have
\begin{eqnarray}
&\frac{T}{1!} \sigma^2
+ \frac{T^2}{2!} (-2\gamma) \sigma^2
+ \frac{T^3}{3!} (-2\gamma)^2 \sigma^2 \cdots \nonumber \\
&= - \frac{\sigma^2}{2\gamma}
\left[ 1 +\frac{T}{1!}(-2\gamma) + \frac{T^2}{2!}(-2\gamma)^2 + \frac{T^3}{3!}(-2\gamma)^3 +  \cdots \right] 
+ \frac{\sigma^2}{2\gamma}
\nonumber \\
&= \frac{\sigma^2}{2\gamma}\left( 1  -  \rme^{-2\gamma T} \right).
\end{eqnarray}

Second, we consider the cases in which Event 2 occurs twice.
Possible cases are depicted in Figure~\ref{fig_OU_combinatorics_second}.
Case A in figure~\ref{fig_OU_combinatorics_second} is easily evaluated as follows:
\begin{eqnarray}
\frac{T^2}{2!}(-2\gamma x_\mathrm{c})(-\gamma x_\mathrm{c}) \nonumber.
\end{eqnarray}
As seen in cases B and C in Figure~\ref{fig_OU_combinatorics_second}, Event 1, which causes no state change, should be adequately inserted.
Using the following three identities,
\begin{eqnarray}
\frac{T^m}{m!}
\left[
(-\lambda_1)^{m-2} + (-\lambda_1)^{m-3}(-\lambda_2) + \cdots + (-\lambda_1) (-\lambda_2)^{m-3} + (-\lambda_2)^{m-2}
\right]
\nonumber \\
=
\frac{T^m}{m!} \left[ (-\lambda_1)^{m-1} - (-\lambda_2)^{m-1}\right] \frac{1}{(-\lambda_1) - (-\lambda_2)},
\end{eqnarray}
\begin{eqnarray}
\frac{\rme^{-\lambda_1 T} -1}{(- \lambda_1)}
= T + \frac{1}{2!} (-\lambda_1) T^2 + \frac{1}{3!} (-\lambda_1)^2 T^3 + \cdots,
\end{eqnarray}
and 
\begin{eqnarray}
\left( - \frac{1}{\lambda_1 - \lambda_2} \right)
\left( - \frac{1}{\lambda_1} \right)
\left( \rme^{-\lambda_1 T} - 1\right)
- 
\left( - \frac{1}{\lambda_1 - \lambda_2} \right)
\left( - \frac{1}{\lambda_2} \right)
\left( \rme^{-\lambda_2 T} - 1\right)  \nonumber \\
=
\int_0^T \rmd t_2 \int_0^{t_2} \rmd t_1 \, \rme^{-\lambda_1 t_1} \rme^{-\lambda_2 (t_2-t_1)},
\end{eqnarray}
we obtain the same equation as in \eref{eq_dual_calculation_for_second_moment}
by setting $\lambda_1 = -2\gamma$ and $\lambda_2 = -\gamma$.
Finally, 
the contribution from Event 2, $(-2\gamma x_\mathrm{c})(-\gamma x_\mathrm{c})$, should be multiplied.

\section{Numerical applications}

As mentioned in the previous section, the expectations of the stochastic differential equation can be evaluated from the discussion on combinatorics for a dual (non-stochastic) process.
Although the Ornstein--Uhlenbeck process is analytically tractable, it is difficult to make such discussions for general cases.
Thus, it is important to evaluate the numerical applicability of the discussions on combinatorics. 
It is difficult to generate all possible paths for high-dimensional cases; however, some combinatorial algorithms are proposed to calculate the expansion coefficients for the exponentials of operators.
In \cite{Amati2019,Zhu2020}, the coefficients required to calculate the Mori--Zwanzig memory kernel were efficiently evaluated by recursive algorithms.
The scenario in this study is essentially similar to that in \cite{Amati2019,Zhu2020}; thus, the following repeated action of $\mathcal{L}^\dagger$ should be evaluated:
\begin{eqnarray}
|n\rangle 
\to \mathcal{L}^\dagger |n\rangle 
\to \left( \mathcal{L}^\dagger \right)^2 |n \rangle
\to \cdots
\to \left( \mathcal{L}^\dagger \right)^M |n \rangle.
\end{eqnarray}
In this study, we only focus on the coefficient for $n=0$ at the final time, because it is sufficient to evaluate the expectations $\mathbb{E}[x - x_\mathrm{c}]$ and $\mathbb{E}[(x - x_\mathrm{c})^2]$ with the initial condition $x = x_\mathrm{c}$ at time $t=0$.
Additionally, we focus on a simple system with only one variable, making it easy to count all possible paths.
For multivariate cases, the discussions made by Zhu et al. \cite{Zhu2020} will be helpful in constructing practical algorithms.

\subsection{Problem settings}

Here, the following one-variable system is considered,
\begin{eqnarray}
\rmd x = - \gamma x^3 \rmd t + \sigma \rmd  W(t),
\label{eq_problem_for_numerical_check}
\end{eqnarray}
which is similar to the Ornstein--Uhlenbeck process, but with a different dependency on $x$ in the drift term.
As in section~\ref{sec_dual_demonstration}, the two expectations, $\mathbb{E}[x - x_\mathrm{c}]$ and $\mathbb{E}[(x - x_\mathrm{c})^2]$, are evaluated in this section.
Therefore, the adjoint operator of the dual process is given as
\begin{eqnarray}
\mathcal{L}^\dagger
= -\gamma x^3 \frac{\partial}{\partial x}
- 3 \gamma x_\mathrm{c} x^2  \frac{\partial}{\partial x}
- 3 \gamma x_\mathrm{c}^2 x  \frac{\partial}{\partial x}
- \gamma x_\mathrm{c}^3  \frac{\partial}{\partial x}
+ \frac{\sigma^2}{2} \frac{\partial^2}{\partial x^2}.
\end{eqnarray}
There are five events:
\begin{enumerate}
\item[(I)] (from $-\gamma x^3 \partial_x$) $n \to n+2$, with a factor $-\gamma n$.
\item[(II)] (from $-3\gamma x_\mathrm{c} x^2 \partial_x$) $n \to n+1$, and with $-3\gamma x_\mathrm{c} n$.
\item[(III)] (from $-3\gamma x_\mathrm{c}^2 x \partial_x$) $n \to n$ (no state change), with a factor $-3\gamma x_\mathrm{c}^2n$.
\item[(IV)] (from $-\gamma x_\mathrm{c}^3 \partial_x$) $n \to n-1$, with a factor $-\gamma x_\mathrm{c}^3 n$.
\item[(V)] (from $(\sigma^2/2) \partial^2_x$) $n \to n-2$, with a factor $\sigma^2 n(n-1)/2$.
\end{enumerate}

\subsection{Taylor series and Pad{\'e} approximation}
\label{sec_result_Pade}

As discussed in section~\ref{sec_dual_process}, the time-evolution operator is formally expressed by the exponential; therefore, the following Taylor series is a primary candidate to perform a numerical evaluation:
\begin{eqnarray}
\rme^{\mathcal{L}^\dagger t} \simeq \sum_{m=0}^M \frac{t^m}{m!} \left( \mathcal{L}^\dagger \right)^m.
\label{eq_operator_in_taylor_series}
\end{eqnarray}
As an example, we here employ the following parameters: $\gamma = 2.0$, $\sigma = 2.0$, and $x_\mathrm{c} = 1.0$.
Let $f_{1 \to 0} (t)$ be the contribution of the state change from $n=1$ to $n=0$.
Then, by evaluating the combinatorics of the possible paths numerically, we obtain 
\begin{eqnarray}
\fl
f_{1 \to 0} (t) \simeq - 2 \, t - 6 \, t^2 + 68 \, t^3 
- 218 \, t^4 - 1653 \, t^5 + 23562.4 \, t^6 + \cdots.
\end{eqnarray}
The coefficients exhibit the behavior of increasing oscillations with sign changes.
Although we expect the coefficients to finally decrease because of the existence of $m!$ in the denominator of \eref{eq_operator_in_taylor_series}, $m$ becomes very large in general, and it is thus impractical to calculate the combinatorics up to such a large $m$.

It is easy to see that the simple summation of the Taylor series does not work; therefore, the Pad{\'e} approximation is used here. The order [$m$/$n$] Pad{\'e} approximation is defined as \cite{Numerical_recipe}
\begin{eqnarray}
f(t) \simeq \frac{P_{m}(t)}{Q_{n}(t)} 
= \frac{p_0 + p_1 t + p_2 t^2 + \cdots + p_m t^m}{1 + q_1 t + q_2 + \cdots + q_n t^n}.
\end{eqnarray}
That is, the function is approximately expressed as a rational function in the Pad{\'e} approximation.

\begin{figure}
\begin{center}
\vskip 5mm
\includegraphics[width=75mm]{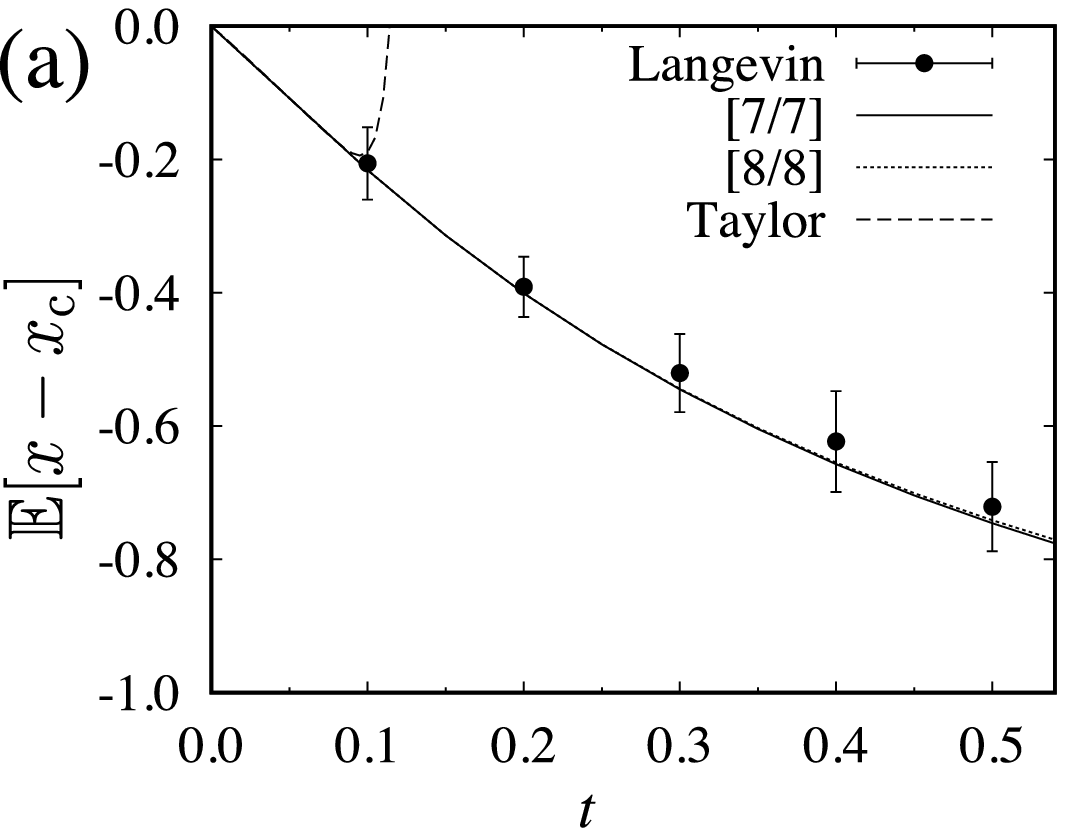}
\hspace{4mm}
\includegraphics[width=75mm]{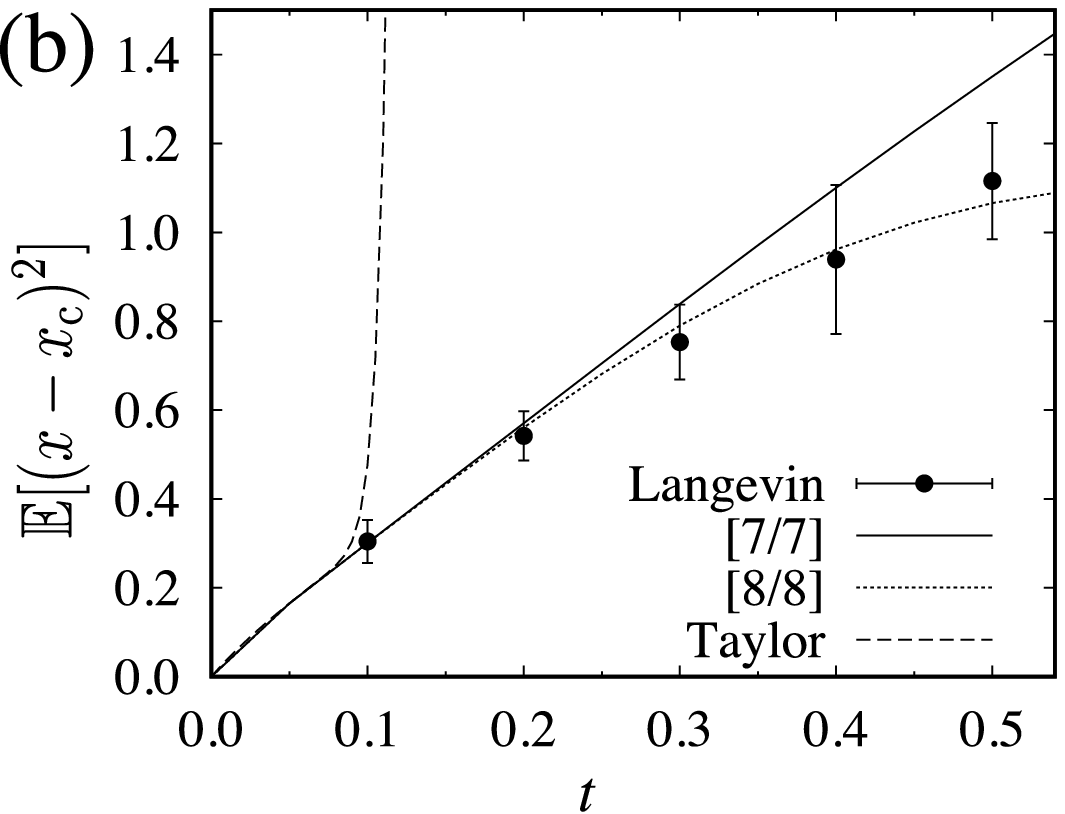}
\end{center}
\caption{Numerical results for \eref{eq_problem_for_numerical_check}. (a) and (b) correspond to expectation values of $\mathbb{E}[x - x_\mathrm{c}]$ and $\mathbb{E}[(x - x_\mathrm{c})^2]$, respectively. Points with error bars denote the results of the Monte Carlo simulations. The solid and dotted lines correspond to [7/7] and [8/8] Pad{\'e} approximations, respectively, and the dashed line is obtained by the simple Taylor series up to the $16$th term.}
\label{fig_result_Pade}
\end{figure}

Figure~\ref{fig_result_Pade} shows the numerical results. 
For comparison, Monte Carlo simulations using the Euler--Maruyama approximation are performed;
further, time-discretization with $10^{-4}$ is employed, and the averages of $1000$ samples are taken.
To depict the error bars, the same simulations with different random-number seeds are performed $10$ times.
Additionally, [7/7] and [8/8] Pad{\'e} approximation results, and those of the simple Taylor summation up to the 16th term, are depicted for $\mathbb{E}[x - x_\mathrm{c}]$ and $\mathbb{E}[(x - x_\mathrm{c})^2]$.
It is clearly seen that the simple Taylor summation exhibits a sudden diverging behavior as $t$ increases.
In contrast, the [8/8] Pad{\'e} approximation provides reasonable results even for the large $t$ cases.

Note that it is not always guaranteed that higher-order approximations yield better results;
the denominator $Q_n(t)$ sometimes takes a small value, which produces unstable results.
However, the following facts are clarified:
\begin{itemize}
\item The simple Taylor summation is not applicable in practical cases.
\item Additional techniques, such as Pad{\'e} approximations, are necessary.
\item For small time-interval cases, the algorithm provides sufficiently accurate estimations. Although the accuracy of the approximations would be insufficient for large time-interval cases, rough estimations are possible.
\end{itemize}

\subsection{Usage of resolvent and Aitken acceleration}
\label{sec_result_resolvent}

In section~\ref{sec_result_Pade}, we employed the following simple Taylor series to interpret the time-evolution operator:
\begin{eqnarray}
\rme^{\mathcal{L}^\dagger t}
= \sum_{m=0}^\infty \frac{t^m}{m!} (\mathcal{L}^\dagger)^m.
\end{eqnarray}
However, it is known that the definition based on the Taylor series is difficult and numerically unusable if $\mathcal{L}^\dagger$ is an unbounded operator in a Banach space \cite{Kato_book}.
Alternatively, the following formula is available:
\begin{eqnarray}
\rme^{\mathcal{L}^\dagger t}
= \lim_{M \to \infty} \left[ \left(1 - \frac{t}{M} \mathcal{L}^\dagger \right)^{-1} \right]^M,
\end{eqnarray}
where $\left(1 - \frac{t}{M} \mathcal{L}^\dagger \right)^{-1}$ is a resolvent of $\mathcal{L}^\dagger$,
apart from a constant factor \cite{Kato_book}.
In this section, this definition for the expansion of the time-evolution operator is employed and investigated.

\begin{figure}
\begin{center}
\vskip 5mm
\includegraphics[width=75mm]{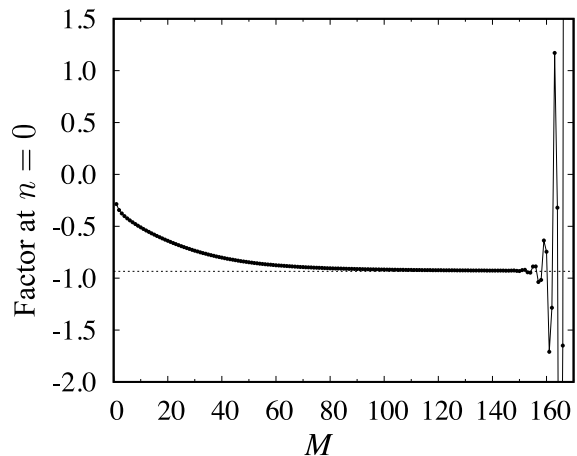}
\end{center}
\caption{Dependence of the factor at $n=0$ on $M$. The parameters $\gamma = 2.0$, $\sigma = 2.0$, $x_\mathrm{c} = 1.0$, and $t=1.0$ are used. As $M$ increases, the factor exhibits convergence behavior; however, a larger $M$ makes the value unstable because of the approximation of the resolvent. The dotted horizontal line corresponds to the mean value of the corresponding Monte Carlo simulations.} 
\label{fig_result_resolvent_pre}
\end{figure}

Because $\mathcal{L}^\dagger$ is expressed as an infinite-dimensional matrix, obtaining the inverse matrix of $1 - \frac{t}{M} \mathcal{L}^\dagger$ is difficult.
However, neglecting rigorous mathematical discussions, the conventional Gauss elimination method is employed here as follows: 
Consider the infinite-dimensional vector 
\begin{eqnarray}
| n \rangle = 
( 0 \quad \cdots \quad 0 \quad 1 \quad 0 \quad \cdots )^\mathrm{T},
\end{eqnarray}
where only the $n$th element is $1$.
The infinite-dimensional matrix corresponding to $1-\frac{t}{M}\mathcal{L}^\dagger$ is then written in the following form:
\begin{eqnarray*}
\left( \begin{array}{ccccc}
\ddots & \vdots       & \vdots  &   \vdots  & \vdots    \\
\cdots & e_{n-1} &    0    &   0      & \cdots    \\
\cdots & d_{n-1} &   e_{n} &   0       & \cdots   \\
\cdots & 1-c_{n-1} & d_{n} &   e_{n+1} & \cdots     \\
\cdots & b_{n-1} &  1-c_{n} &  d_{n+1} & \cdots \\
\cdots & a_{n-1} &   b_{n} &   1-c_{n+1} &  \cdots\\
\cdots & 0      &   a_{n} &   b_{n+1} & \cdots \\
\cdots & 0      &   0  &      a_{n+1} & \cdots \\
\vdots & \vdots     &   \vdots  &    \vdots & \ddots \\
\end{array} \right),
\end{eqnarray*}
where $a_n, \dots, e_n$ correspond to factors of the above-mentioned five events, respectively.
Note that each factor is multiplied by $t/M$.
Therefore, if $M$ is sufficiently large,
the $n$th column in the inverse matrix, $\left(1 - \frac{t}{M} \mathcal{L}^\dagger \right)^{-1}$, can be approximately expressed as follows:
\begin{eqnarray*}
\begin{array}{c}
    \vdots   \\
   0    \\
  (-e_{n})(1-c_{n-2})^{-1} (1-c_{n})^{-1}    \\
 (-d_{n})(1-c_{n-1})^{-1} (1-c_{n})^{-1}   \\
 (1-c_{n})^{-1}   \\
  (-b_{n})(1-c_{n+1})^{-1} (1-c_{n})^{-1}   \\
 (-a_{n})(1-c_{n+2})^{-1} (1-c_{n})^{-1}  \\
   0   \\
 \vdots
\end{array}.
\end{eqnarray*}
That is, if $M$ is large, $a_n, \dots, e_n$ are sufficiently small; then, for example, $b_{n} e_{n+1}$ could be sufficiently small to be neglected.
Thus, using the combinatorial algorithm, it is possible to evaluate 
\begin{eqnarray}
\rme^{\mathcal{L}^\dagger t} |n\rangle
\simeq \left[ \left(1 - \frac{t}{M} \mathcal{L}^\dagger \right)^{-1} \right]^M |n\rangle
\label{eq_repeated_resolvent}
\end{eqnarray}
approximately.
At the final step, the factor at state $n=0$ corresponds to the target statistics.

To evaluate $\mathbb{E}[x - x_\mathrm{c}]$, the initial state is set to $n=1$. 
The dependence of the factor at $n=0$ on $M$ is depicted in Figure~\ref{fig_result_resolvent_pre}, in which time $t = 1.0$, and the same parameters as those in section~\ref{sec_result_Pade}, i.e., $\gamma = 2.0$, $\sigma = 2.0$, and $x_\mathrm{c} = 1.0$, are used.
As $M$ increases, the factor appears to converge to a certain value.
However, a larger $M$ causes unstable and diverging behavior, as shown in the region $M \ge 150$ in Figure~\ref{fig_result_resolvent_pre}.
The resolvent used here is an approximate one, which can cause diverging behavior.

\begin{figure}
\begin{center}
\vskip 5mm
\includegraphics[width=75mm]{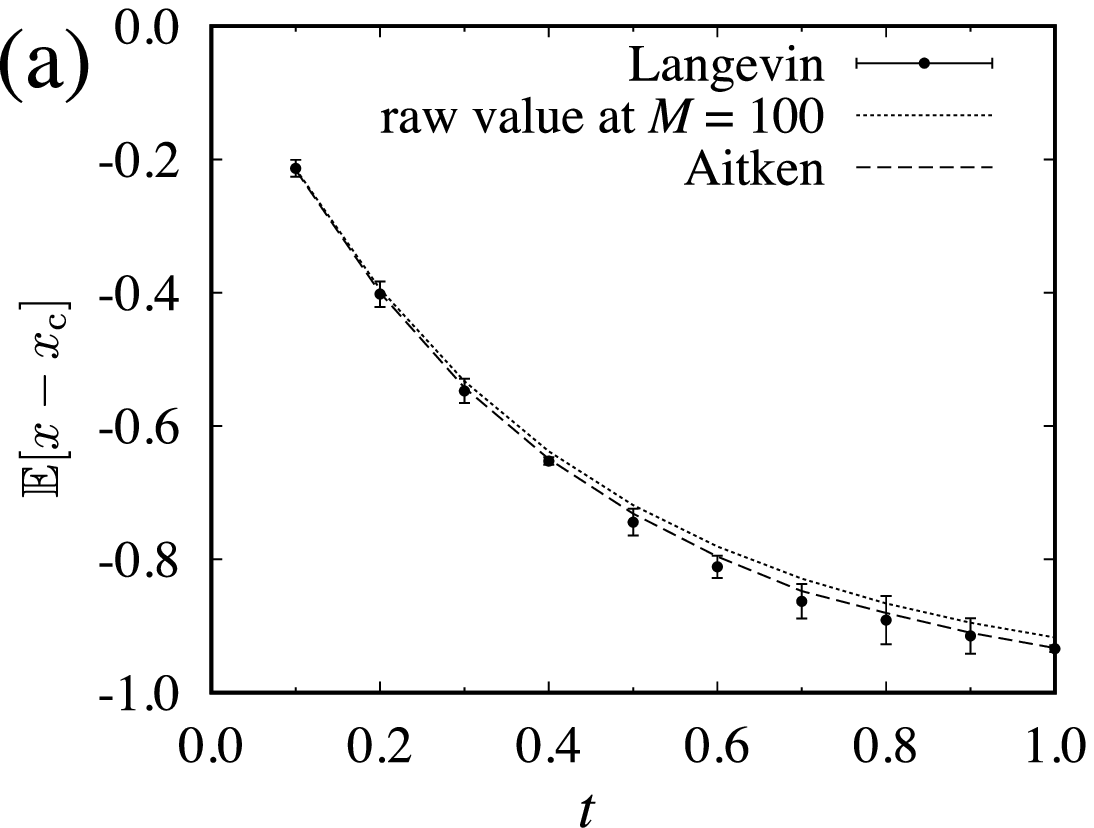}
\hspace{4mm}
\includegraphics[width=75mm]{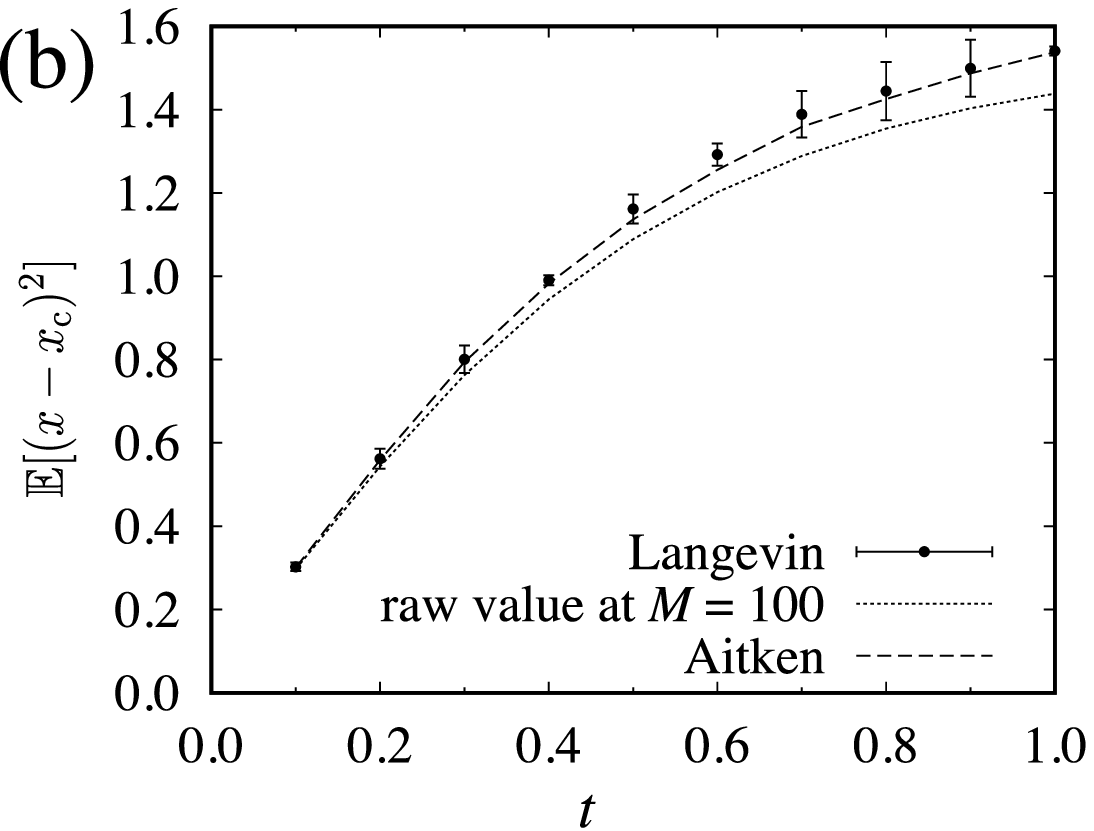}
\end{center}
\caption{Numerical results obtained using the resolvent in \eref{eq_repeated_resolvent}. (a) and (b) correspond to the expectation values of $\mathbb{E}[x - x_\mathrm{c}]$ and $\mathbb{E}[(x - x_\mathrm{c})^2]$, respectively. Points with error bars are results of the Monte Carlo simulations. The dotted line corresponds to the raw value of the factor at $M=100$. The dashed line is obtained by the Aitken acceleration method using the sequence from $M=0$ to $M=100$.}
\label{fig_result_resolvent_moments}
\end{figure}

To avoid this unstable numerical behavior, we consider the use of only the sequences in the stable regions. 
A simple way is to find the convergent point from Figure~\ref{fig_result_resolvent_pre}. 
However, if we change the parameters, e.g., time $t$, the diverging behavior becomes different from that in Figure~\ref{fig_result_resolvent_pre}. 
This detailed analysis is cumbersome; therefore, to avoid this complex analysis, we apply the Aitken acceleration method using only the series in the stable region, as follows \cite{Numerical_recipe}:
\begin{eqnarray}
z^{(k+1)}_m = z^{(k)}_m - \frac{(z^{(k)}_{m+1} - z^{(k)}_m)^2} {z^{(k)}_{m+2} - 2 z^{(k)}_{m+1} + z^{(k)}_m},
\end{eqnarray}
here, the series from $M = 0$ to $M=100$ was used.
The Aitken acceleration is repeatedly applied 10 times, i.e., $k = 10$ cases are evaluated.
The evaluated results for $\mathbb{E}[x - x_\mathrm{c}]$ and $\mathbb{E}[(x - x_\mathrm{c})^2]$ are shown in Figure~\ref{fig_result_resolvent_moments}.
Even for large $t$ regions, the Aitken acceleration method works well.

\section{Concluding remarks}

In this study, a discussion on combinatorics for calculating the expectations of stochastic differential equations was presented. 
The discussion naturally connects the duality relations in stochastic processes and combinatorics. 
The combinatorics was effective in recovering the established results of the Ornstein--Uhlenbeck process.
Furthermore, it is important to find numerical methods for the practical use of combinatorics; hence, two candidates were proposed.
One is the use of the Pad{\'e} approximation in the conventional Taylor-type expansion of the time-evolution operator.
The other is based on the resolvent of the time-evolution operator, with the application of Aitken acceleration.
Both these methods produce reasonable approximations; in particular, the use of the resolvent and Aitken acceleration seems to work well.
Note that numerical verifications were performed only for limited cases. Therefore, further studies involving higher-dimensional cases should be performed in the future.

The usage of the adjoint description and duality relations is natural for the following reason: Our aim here is not to calculate the probability density function but to calculate the expectations. If we use a simple expansion for the original continuous partial differential equation (Fokker-Planck equation), it needs expansion with considerably high degrees. In addition,  it is difficult to describe the initial conditions of the Dirac delta function in such expansions. However, using the adjoint description, the degree of the expansion directly corresponds to the statistics we want to evaluate, and it can also avoid the initial condition problem. Hence, the method proposed in this study is fit for our aim.

Finally, we note the following points.
Mathematically rigorous discussions should be performed in future works, especially for using an approximated resolvent and its complicated divergent behavior.
The repeated applications of the resolvent in \eref{eq_repeated_resolvent} resemble a simple Bernoulli trial; its limit with $M \to \infty$ could lead to a conventional Poisson process.
This consideration might be important in constructing more stable numerical algorithms.
Additionally, in studies on time-evolution equations, it is common to employ numerical time-integration algorithms with a discrete time interval $\Delta t$, and it is possible to discuss the approximation errors for the order of $\Delta t$.
In contrast, the proposed methods do not use such time-discretization directly.
As for the usage of Pad{\'e} approximation, an approximation by a rational function of a given order is obtained, and it gives useful information when trying to discuss functional forms of statistics.
In section 4.3, the truncation of summations or finite-state approximations of matrices of infinite sizes is performed.
The discussion in this paper gives us methods to evaluate the expectation values directly without using Monte Carlo samplings and evaluating the probability density functions.
Of course, numerical studies on such algorithms are just beginning to be conducted; discussions regarding the approximation errors remain to be explored in future works.
However, such different concepts of time-integration can aid future works in efficient calculations of expectation values.

\ack
This work was supported by JST, PRESTO Grant Number JPMJPR18M4, Japan.\\


\vspace{3mm}


\begin{thebibliography}{99}

\bibitem{Gardiner_book}
Gardiner C 2009
{\it Stochastic methods: A handbook for the natural and social sciences, 4th edition}.
(Berlin Heidelberg: Springer)


\bibitem{Kloeden_book}
Kloeden P E and Platen E 1992 {\it Numerical Solution of Stochastic Differential Equations}
(Berlin: Springer)


\bibitem{Liggett_book} Liggett T M 2005 
{\it Interacting Particle Systems (Classics in Mathematics)} (Berlin: Springer)
Reprint of the 1985 edition

\bibitem{Jansen2014}
Jansen S and Kurt N 2014 {\it Probab. Surveys} {\bf 11} 59


\bibitem{Shiga1986} Shiga T and Uchiyama K 1986 {\it Probab. Th. Rel. Fields} {\bf 73} 87
\bibitem{Mohle1999} M{\"o}hle M 1999 {\it Bernoulli} {\bf 5} 761
\bibitem{Carinci2015} Carinci G, Giardin{\`a} C, Giberti C and Redig F 2015 {\it Stochastic Processes and their Applications} {\bf 125} 941

\bibitem{Giardina2007} Giardin{\`a} C, Kurchan J and Redig F 2007 {\it J. Math. Phys.} {\bf 48} 033301
\bibitem{Carinci2013}
Carinci G, Giardin{\'a} C, Giberti C and Redig F 2013 {\it J. Stat. Phys.} {\bf 152} 657


\bibitem{Giardina2009} Giardin{\`a} C, Kurchan J, Redig F and Vafayi K 2009 {\it J. Stat. Phys.} {\bf 135} 25
\bibitem{Franceschini2017} Franceschini C and Giardin{\`a} C 2019 {\it Sojourns in Probability Theory and Statistical Physics - III}  ed V Sidoravicius (Singapore: Springer) pp 187-214

\bibitem{Redig2018} Redig F and Sau F 2018 {\it J. Stat. Phys.} {\bf 172} 980
\bibitem{Groenevelt2019} Groenevelt W 2019 {\it J. Stat. Phys.} {\bf 174} 97


\bibitem{Ohkubo2015} Ohkubo J 2015 {\it Phys. Rev. E} {\bf 92} 043302

\bibitem{Amati2019}
Amati G, Meyer H and Schilling T 2019 {\it J. Stat. Phys.} {\bf 174} 219
\bibitem{Zhu2020}
Zhu Y and Venturi D 2020 {\it J. Stat. Phys.} {\bf 178} 1217

\bibitem{Schutz1997} Sch{\"u}tz G M 1997 {\it J. Stat. Phys.} {\bf 86} 1265
\bibitem{Imamura2011} Imamura T and Sasamoto T 2011 {\it J. Stat. Phys.} {\bf 142} 919
\bibitem{Ohkubo2017} Ohkubo J 2017 {\it J. Phys. A: Math. Gen.} {\bf 50} 095004


\bibitem{Ohkubo2019}
Ohkubo J and Arai Y 2019 {\it J. Stat. Mech.} 063202

\bibitem{Ohkubo2020}
Ohkubo J 2020 {\it J. Phys. Soc. Jpn.} {\bf 89} 044004


\bibitem{Ohkubo2013} Ohkubo J 2013 {\it J. Phys. A: Math. Theor.} {\bf 46} 375004


\bibitem{Gillespie1977}
Gillespie D T 1977 {\it J. Phys. Chem.} {\bf 81} 2340


\bibitem{Wilkinson_book}
Wilkinson D J 2006
{\it Stochastic Modelling for Systems Biology}
(Boca Raton: CRC Press)



\bibitem{Warne2019}
Warne D J, Baker R E and Simpson M J 2019 {\it J. R. Soc. Interface} {\bf 16} 20180943


\bibitem{Ohkubo2010} Ohkubo J 2010 {\it J. Stat. Phys.} {\bf 139} 454


\bibitem{Weber2017}
Weber M F and Frey E 2017 {\it Rep. Prog. Phys.} {\bf 80} 046601



\bibitem{Kleinert_book} 
Kleinert H 2006
{\it Path Integrals in Quantum Mechanics, Statistics, Polymer Physics, and Financial Markets, 4th Edition}
(Danvers: World Scientific Publishing)


\bibitem{Voliotis2016}
Voliotis M, Thomas P, Grima R and Bowsher C G 2016 {\it PLoS Comput. Biol.} {\bf 12} e1004923

\bibitem{Beentjes2019}
Beentjes C H L and Baker R E 2019 {\it J. Chem. Phys.} {\bf 150} 154107


\bibitem{Numerical_recipe}
Press W H, Teukolsky S A, Vetterling W T and Flannery B P
{\it Numerical Recipes in C, 2nd edition}
(Cambridge: Cambridge University Press)


\bibitem{Kato_book}
Kato T 1966
{\it Perturbation Theory for Linear Operators}
(Berlin: Springer)







\end{thebibliography}
\end{document}